\documentclass{PoS}

\newcommand{\bea}{\begin{eqnarray}}
\newcommand{\eea}{\end{eqnarray}}
\newcommand{\beq}{\begin{equation}}
\newcommand{\eeq}{\end{equation}}
\newcommand{\nn}{\nonumber}
\newcommand{\gev}{{\rm GeV}}
\newcommand{\mev}{{\rm MeV}}
\newcommand{\msb}{\overline{\rm{MS}}}

\def\dfrac#1#2{{\displaystyle {#1 \over #2}}}

\def\simge{\mathrel{\rlap{\raise 0.511ex \hbox{$>$}}{\lower 0.511ex
 \hbox{$\sim$}}}}
\def\simle{\mathrel{\rlap{\raise 0.511ex \hbox{$<$}}{\lower 0.511ex
 \hbox{$\sim$}}}}
\def\slash#1{\setbox0=\hbox{$#1$}\dimen0=\wd0 \setbox1=\hbox{/} \dimen1=\wd1
 \ifdim\dimen0>\dimen1 \rlap{\hbox to \dimen0{\hfil/\hfil}} #1
 \else \rlap{\hbox to \dimen1{\hfil$#1$\hfil}} / \fi}

\title{Light quark masses and pseudoscalar decay constants from $N_f=2$
  twisted mass QCD}

\ShortTitle{Light quark masses and pseudoscalar decay constants from $N_f=2$
  tmQCD}

\author{Vittorio Lubicz\\
        Dip. di Fisica, Universit{\`a} di Roma Tre and INFN, Sez. di Roma
Tre,\\ Via della Vasca Navale 84, I-00146 Roma, Italy\\
        E-mail: \email{lubicz@fis.uniroma3.it}}

\author{Silvano Simula\\
        INFN, Sez. di Roma Tre,\\ 
Via della Vasca Navale 84, I-00146 Roma, Italy\\
        E-mail: \email{silvano.simula@roma3.infn.it}}

\author{\speaker{Cecilia Tarantino}\thanks{It is a pleasure to thank the
        organizers of ``Lattice 2007'' for the very interesting
        conference realized in Regensburg. We  
        thank the other authors of the work presented here:
        B.~Blossier, Ph.~Boucaud, P.~Dimopoulos, F.~Farchioni, R.~Frezzotti, 
V.~Gimenez, G.~Herdoiza, K.~Jansen, C.~Michael, D.~Palao, M.~Papinutto,
A.~Shindler, C.~Urbach, and U.~Wenger . We are also grateful to
        D.~Becirevic, G.~Martinelli and G.C.~Rossi for useful comments and
discussions.}\\
        Dip. di Fisica, Universit{\`a} di Roma Tre and INFN, Sez. di Roma
Tre,\\ Via della Vasca Navale 84, I-00146 Roma, Italy\\
        E-mail: \email{tarantino@fis.uniroma3.it}}

\author{for the European Twisted Mass Collaboration (ETMC)}

\abstract{We present the results of the lattice QCD
  calculation of the average up-down and
strange quark masses and of the light meson pseudoscalar decay constants,
  recently performed with $N_f=2$ dynamical fermions by the ETM Collaboration. The simulation is carried out at a single value of
the lattice spacing with the twisted mass fermionic action at maximal twist,
which guarantees automatic ${\cal O}(a)$-improvement of the physical quantities.
Quark masses are renormalized by implementing the non perturbative RI-MOM
renormalization procedure. Our results for the light quark masses are
$m_{ud}^{\msb}(2\ \gev)=3.85 \pm 0.12 \pm 0.40$ MeV, $m_{s}^{\msb}(2\ \gev)=105
\pm 3 \pm 9$ MeV and $m_s/m_{ud}=27.3 \pm 0.3 \pm 1.2$. We also obtain
$f_K=161.7 \pm 1.2 \pm 3.1$ MeV and the ratio $f_K/f_\pi=1.227 \pm 0.009  \pm
0.024$. From this ratio, by using the experimental determination of $\Gamma(K
\to \mu \bar \nu_\mu (\gamma))/\Gamma(\pi \to \mu \bar \nu_\mu (\gamma))$ and
the average value of $\vert V_{ud}\vert$ from nuclear beta decays, we obtain
$\vert V_{us}\vert=0.2192(5)(45)$, in agreement with the determination from
$K_{l3}$ decays and the unitarity constraint.}

\FullConference{The XXV International Symposium on Lattice Field Theory\\
                 July 30-4 August 2007\\
                 Regensburg, Germany}

\begin{document}
%%%%%%%%%%%%%%%%%%%%%%%%%%%%%%%%%%%%%%%%%%%%%%%%%%%%%%%%%%%%%%%%%%%%%
\begin{figure}[h]
\begin{center}
\vspace*{-0.6cm}
\includegraphics[scale=0.28,angle=270]{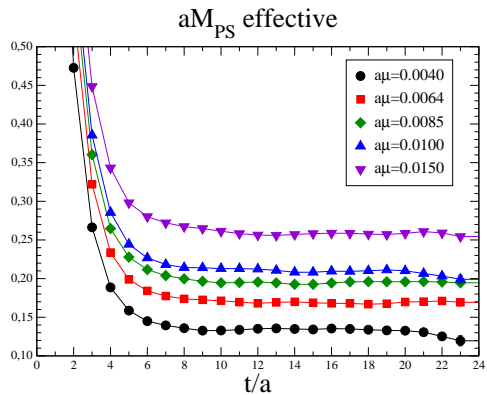}
\vspace*{-0.6cm}
\caption{\label{fig:effmass} \small\sl Effective masses of pseudoscalar mesons
  with $\mu_S=\mu_1=\mu_2$, as a function of the
time.}
\end{center}
\end{figure}
%%%%%%%%%%%%%%%%%%%%%%%%%%%%%%%%%%%%%%%%%%%%%%%%%%%%%%%%%%%%%%%%%%%%%
\vspace*{-0.9cm}
\section{Introduction}
\label{sec:intro}
We present our recent determination~\cite{NOI}
of the light quark masses (strange quark mass $m_s$ and average up-down
quark mass $m_{ud}$), of the kaon pseudoscalar decay constant $f_K$, and of the
ratio $f_K/f_\pi$.  In order to investigate
the properties of the $K$ meson, we have simulated the
theory with $N_f=2$ degenerate dynamical and two valence 
quarks, considering  a partially
quenched setup with the valence quark masses $\mu_1$ and $\mu_2$
different between each other and from the sea quark mass
$\mu_S$.

The calculation is based on a set of (ETMC) gauge field configurations
generated with the tree-level improved Symanzik gauge action at $\beta=3.9$, corresponding to
$a=0.087(1)$ fm ($a^{-1} \simeq 2.3$ GeV)~\cite{Boucaud:2007uk}, and the
twisted mass fermionic action at maximal twist. We have simulated 5 values of
the bare sea quark mass,
$a\mu_S=\{0.0040,\, 0.0064,\, 0.0085,\, 0.0100,\, 0.0150\}$
and 8 values, $a\mu_{1,2}=\{0.0040,\, 0.0064,\, 0.0085,\, 0.0100,\, 0.0150,\, 0.0220,\,
0.0270,\, 0.0320\}$, for the valence quark mass.
The first five masses, equal to the sea masses, lie in the range
$1/6\, m_s \simle \mu_{1,2} \simle 2/3\, m_s$, being $m_s$ the physical
strange quark mass, while the heaviest three are around the strange quark mass.

At each value of the sea quark mass we have computed the two-point correlation
functions of charged pseudoscalar mesons, on a set of 240 independent gauge
field configurations, separated by 20 HMC trajectories one from the other. 
To improve
the statistical accuracy, we have evaluated the meson correlators using a
stochastic method with a $Z(2)$-noise to include all spatial sources~\cite{Foster:1998vw,McNeile:2006bz}.
Statistical errors on meson masses and decay constants are evaluated using
the jackknife procedure, while those on the fit results, based on
data obtained at different sea quark masses, are evaluated using a bootstrap
procedure. Further details on the numerical simulation can be found in
refs.~\cite{NOI,Urbach}.

The use of twisted mass fermions presents several advantages~\cite{Frezzotti:2003ni}: i) the pseudoscalar meson
masses and decay constants are automatically improved at ${\cal O}(a)$; ii) at
maximal twist, the physical quark mass is directly related to the twisted mass
parameter of the action, and it is subject only to multiplicative
renormalization; iii) the determination of the pseudoscalar decay constant does
not require the introduction of any renormalization constant, and it is based on
the relation
\beq
f_{PS} = (\mu_1 + \mu_2) \frac{\vert \langle 0 \vert P^1(0) \vert P \rangle\vert}
{M_{PS}^{\,2}}\, .
\eeq
The meson mass $M_{PS}$ and the matrix element $\vert\langle 0 \vert P^1(0)
\vert P \rangle\vert$ have been extracted from a fit of the two-point
pseudoscalar correlation function in the time interval $t/a \in [10,21]$. In
order to illustrate the quality of the data, we show in fig.~\ref{fig:effmass}
the effective masses of pseudoscalar mesons, as a function of the time, in the
degenerate cases $\mu_S=\mu_1=\mu_2$.

\section{Quark mass dependence of pseudoscalar meson mas\-ses and decay
constants}
The determination of the physical properties of $K$ mesons requires to study 
the corresponding observables over a large range of
masses, from the physical strange quark down to the light up-down
quark. In ref.~\cite{NOI}, we have studied the quark mass dependence of
pseudoscalar meson masses and decay constants by considering two different functional
forms: i) the dependence predicted by continuum partially quenched
chiral perturbation theory (PQChPT), ii) a polynomial dependence. 

\underline{\bf PQChPT fits:}
Within PQChPT we have considered the full next-to-leading order (NLO) expressions with
the addition of the local NNLO contributions, i.e. terms quadratic in the quark
masses, which turn out to be needed for a good description of the data up to the region of the strange quark. The PQChPT
predictions~\cite{sharpe} can be written as
\bea
\label{eq:mf12}
&& M_{PS}^{\,2}(\mu_S,\mu_1,\mu_2) \ =\ B_0 \left(\mu_1 + \mu_2\right) \cdot
\left[ 1 + \dfrac{\xi_1\, (\xi_S-\xi_1)\, \ln 2 \xi_1}{(\xi_2-\xi_1)} -
\dfrac{\xi_2\, (\xi_S-\xi_2)\, \ln 2 \xi_2}{(\xi_2-\xi_1)} + \nn \right.\\
&& \qquad
\left. + a_V\, \xi_{12} + a_S\, \xi_S + a_{VV}\, \xi_{12}^2 + a_{SS}\, \xi_S^2 
+ a_{VS}\, \xi_{12}\, \xi_S + a_{VD}\, \xi_{D12}^2 \right] \, , \\
&& f_{PS}(\mu_S,\mu_1,\mu_2) \ =\ f \cdot
\left[ 1 - \xi_{1S}\, \ln 2 \xi_{1S} - \xi_{2S}\, \ln 2 \xi_{2S} +
\dfrac{\xi_1\,\xi_2 - \xi_S\,\xi_{12}}{2\,(\xi_2-\xi_1)} \ln \left( 
\frac{\xi_1}{\xi_2}\right) + \nn \right.\\
&& \quad \left. + (b_V + 1/2) \, \xi_{12} + (b_S - 1/2)\, \xi_S + b_{VV}\,
\xi_{12}^2 + b_{SS}\, \xi_S^2 + b_{VS}\, \xi_{12}\, \xi_S + b_{VD}\, \xi_{D12}^2
\right]\, , \nn
\eea
where $\xi_i =2 B_0 \mu_i/(4\pi f)^2$, $\xi_{ij}=B_0 (\mu_i+\mu_j)/(4\pi f)^2$
and $\xi_{Dij}=B_0 (\mu_i-\mu_j)/(4\pi f)^2$. The parameters $B_0$ and $f$ are
the LO low energy constants (LECs)~\footnote{The pseudoscalar decay constant $f$ is normalised such that $f_\pi
= 130.7$ MeV at the physical pion mass.}, whereas $a_V$, $a_S$, $b_V$ and $b_S$
are related to the NLO LECs by
$a_V = 4 \alpha_8 - 2 \alpha_5$, $a_S = 8 \alpha_6 - 4 \alpha_4$, $b_V = \alpha_5$, $b_S = 2  \alpha_4$.
The quadratic mass terms in eq.~(\ref{eq:mf12}) represent the
local NNLO contributions. The chiral logarithms, also known at two loops in
the partially quenched theory~\cite{Bijnens:2005pa},
involve a larger number of NLO LECs whose values cannot be fixed from
phenomenology in the $N_f=2$ theory. Introducing their contribution 
would increase significantly the number of free parameters, thus
limiting the predictive power of the calculation.

Aiming at a percent precision, the impact of finite size corrections cannot be
neglected in our study, where the
lattice spatial extension is $L=24 a \simeq 2.2$ fm and 
$M_{PS} L \geq 3.2$. Since we have not performed yet a systematic study on 
different lattice volumes,
we have estimated the finite size effects by including in the fits the 
corrections predicted by one-loop PQChPT~\cite{bv} (for their explicit
expressions see ref.~\cite{NOI}).

\underline{\bf Polynomial fits:}
The inclusion of the local NNLO contributions in the PQChPT predictions
of eq.~(\ref{eq:mf12}) is required by the observation that the pure
NLO predictions are not accurate enough to describe the quark mass dependence of
pseudoscalar meson masses and decay constants up to the strange quark region. Not having considered the full NNLO chiral predictions, we
have  evaluated the associated
systematic uncertainty, considering as an alternative
description a simple polynomial dependence on the quark masses, for
both pseudoscalar meson masses and decay constants:
\bea
\label{eq:mf12pol}
&M_{PS}^{\,2}(\mu_S,\mu_1,\mu_2) &= B_0\left(\mu_1 + \mu_2\right)\cdot
\left[ 1 +  a_V \xi_{12} + a_S \xi_S + a_{VV} \xi_{12}^2 + a_{SS}
\xi_S^2 + a_{VS} \xi_{12} \xi_S + a_{VD} \xi_{D12}^2 \right],\nn\\
&f_{PS}(\mu_S,\mu_1,\mu_2) &= f \cdot
\left[ 1  + (b_V + 1/2)  \xi_{12} + (b_S - 1/2) \xi_S + b_{VV} \xi_{12}^2
+ b_{SS} \xi_S^2 +  b_{VS} \xi_{12} \xi_S + b_{VD} \xi_{D12}^2
\right]. \nn\\ \hspace*{-3.0cm}
\eea
The differences between the results obtained by performing either chiral
or polynomial fits have been included in the final estimates of the systematic
errors.

\section{Chiral extrapolations}
The input data of our analysis~\cite{NOI} are the lattice results for the
pseudoscalar meson masses and decay constants obtained at each value of the sea
quark mass, with both degenerate and non degenerate valence quarks. We have
excluded from the fits the heaviest mesons having both the valence quark masses in the
strange mass region, i.e. with $a \mu_{1,2}=\{0.0220,\, 0.0270,\, 0.0320\}$,
considering therefore 150 combinations of quark masses. 
The number of free
parameters in the combined fit of $M_{PS}^{\,2}$ and $f_{PS}$ is 14, but a first
analysis shows that some of them (from $1$ to $5$ depending on the fit) are compatible with zero
within one standard deviation, and are kept fixed to zero.

In order to extrapolate the pseudoscalar meson masses and decay constants to the
points corresponding to the physical pion and kaon, we have considered
three different fits:
\begin{itemize}
\vspace*{-0.2cm}
\item {\bf\underline{Polynomial fit}:} a polynomial dependence on the quark
masses is assumed for pseudoscalar meson masses and decay constants,
according to eq.~(\ref{eq:mf12pol}).
\vspace*{-0.3cm}
\item {\bf\underline{PQChPT fit}:} pseudoscalar meson masses and decay
constants are fitted according to the PQChPT predictions of
eq.~(\ref{eq:mf12}) including the finite volume corrections derived in ref.~\cite{bv}.
\vspace*{-0.3cm}
\item {\bf\underline{Constrained PQChPT fit}:} this fit, denoted as C-PQChPT in
the following, deserves a more detailed explanation. The main uncertainty in
using eqs.~(\ref{eq:mf12}) and (\ref{eq:mf12pol}) to describe the
quark mass dependence of $M_{PS}^{\,2}$ and $f_{PS}$ is related to the
extrapolation toward the physical up-down quark mass. On the other hand,
we have shown in ref.~\cite{Boucaud:2007uk} that pure NLO ChPT, with the
inclusion of finite volume corrections, is sufficiently accurate in describing
the lattice pseudoscalar meson masses and decay constants
when the analysis is restricted to our lightest four quark masses in the unitary
setup (i.e. $\mu_1=\mu_2=\mu_S$). In order to take advantage of this
information, when performing the C-PQChPT fit we first determine the LO
parameters $B_0$ and $f$ and the NLO combinations $a_V+a_S$ and $b_V+b_S$ from a
fit based on pure NLO ChPT performed on the lightest four unitary points.
By using these constraints, the other parameters entering the chiral 
expansions of
$M_{PS}^{\,2}$ and $f_{PS}$ are then obtained from a fit to eq.~(\ref{eq:mf12})
over the non unitary points. For consistency with the previous unitary fit, we
exclude also in this case from the analysis the data at the highest value of sea
quark mass, $a\mu_S=0.0150$.
\end{itemize}
\vspace*{-0.2cm}
We find that, though the quality of the fit is better
in the polynomial case, all three analyses provide a good description of the
lattice data, in the whole region of masses explored in the simulation, once
the terms quadratic in the quark masses are taken into account.

A potential problem in the partially quenched theory is the divergence of the
chiral logarithms in the limit in which the light valence quark mass goes to
zero at fixed sea quark mass (see eq.~(\ref{eq:mf12})). This divergence does not
affect the extrapolation of the lattice results to the physical point, since 
sea and the light valence quark masses are degenerate in this case. However, in
order to verify that this unphysical behaviour of the partially quenched chiral
logarithms does not modify the result of the extrapolation, we have repeated the
analysis restricting both the polynomial and the chiral fits to the $30$
quark mass combinations (26 in the case of the C-PQChPT fit) that, satisfying
the constraint $\mu_2 \ge \mu_1 = \mu_S$, are not affected by dangerous
chiral logarithms.  The comparison
between the results obtained by considering the two different sets of quark
masses is reassuring, as it shows that the effects of potentially divergent
chiral logarithms are well under control in our analysis.

\begin{figure}[!t]
\vspace{-0.8cm}
\begin{minipage}{16pc}
\includegraphics[scale=0.3,angle=270]{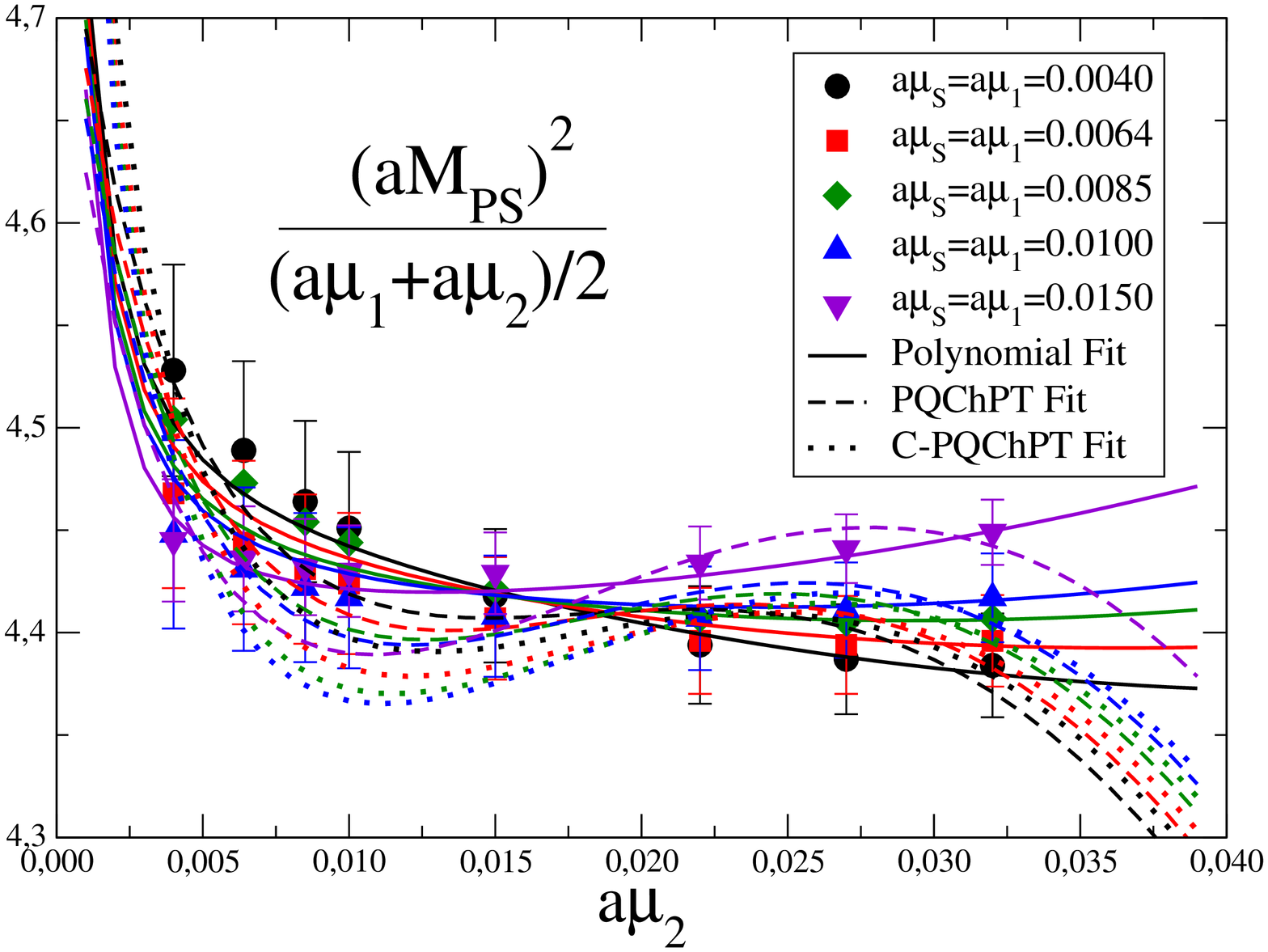}
\end{minipage}\hspace{2pc}
\begin{minipage}{16pc}
\includegraphics[scale=0.3,angle=270]{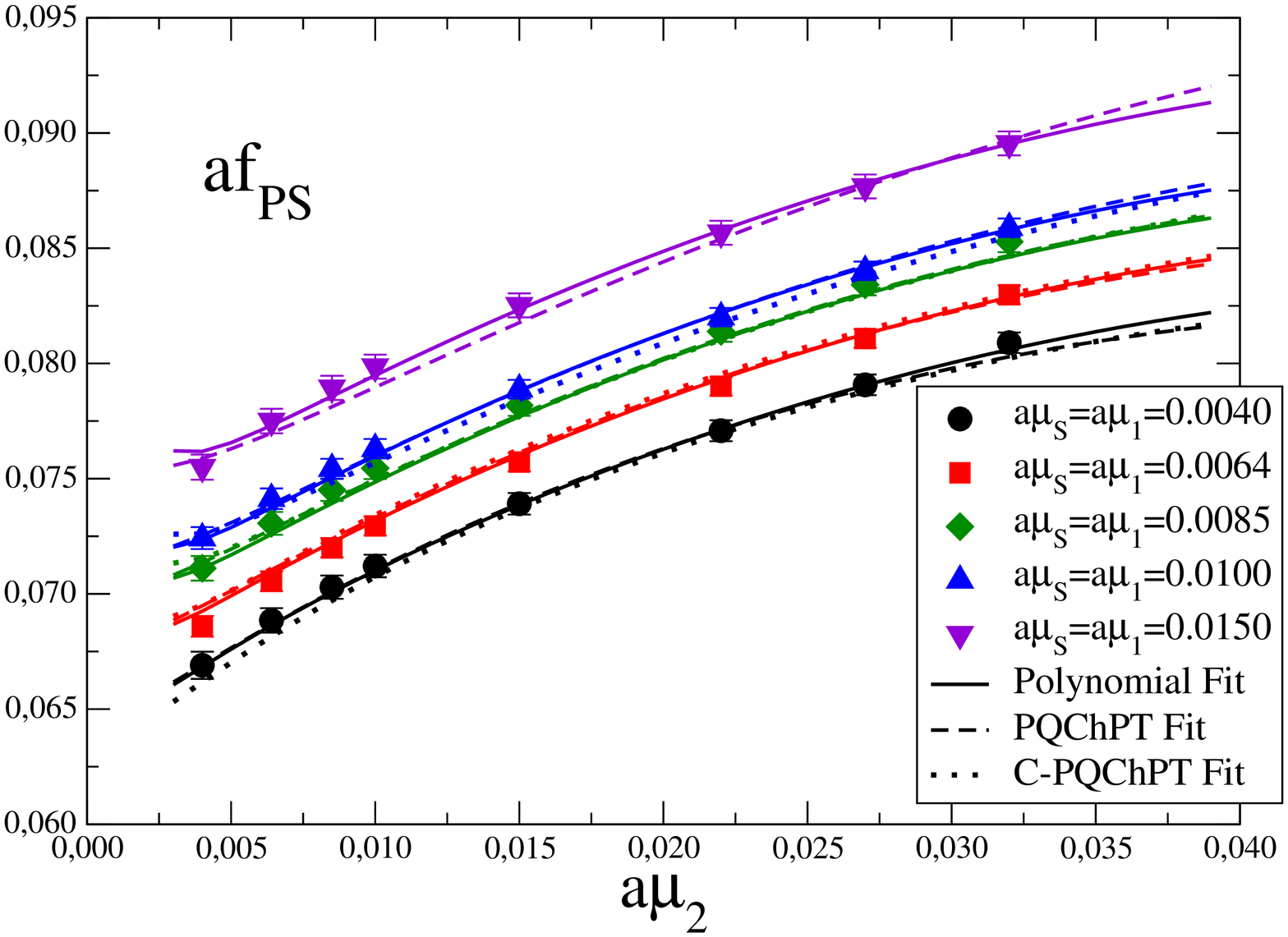}
\end{minipage}
\vspace*{-0.6cm}
\caption{\sl Lattice results for  $a^2 M_{PS}^{\,2}/
\frac{1}{2}(a\mu_1+a\mu_2)$ (left) and $a f_{PS}$ (right) as a function of
the valence quark mass $a\mu_2$, with $a\mu_1=a\mu_S$. The solid, dashed and
dotted curves represent the results of the three fits.}
\label{fig:3fits}
\vspace*{-0.3cm}
\end{figure}

The mass dependence of the pseudoscalar meson masses and decay constants is
illustrated in fig.~\ref{fig:3fits},
where lattice data are compared with the results of the polynomial,
PQChPT and C-PQChPT fits. We have shown in the plots the cases in which one of
the valence quark mass ($\mu_1$) is equal to the sea quark mass, and the
results are presented as a function of the second valence quark mass
($\mu_2$). The points corresponding to the physical pion and kaon are thus
obtained by extrapolating/interpolating the results shown in 
fig.~\ref{fig:3fits} to the limits $\mu_1 \to m_{ud}$ and $\mu_2 \to m_s$.
%%%%%%%%%%%%%%%%%%%%%%%%%%%%%%%%%%%%%%%%%%%%%%%%%%%%%%%%%%%%%%%%%%%%%
%%%%%%%%%%%%%%%%%%%%%%%%%%%%%%%%%%%%%%%%%%%%%%%%%%%%%%%%%%%%%%%%%%%%%

To investigate the impact of finite volume corrections we have compared, for
the pseudoscalar meson masses and decay constants, the
PQChPT fits obtained with or without including these
corrections. The differences turn out to be small~\cite{NOI}; however, to better quantify the systematic error due to finite size effects, we plan to
extend our analysis
on lattices with different spatial sizes.

By having determined the fit parameters, we have then extrapolated
eqs.~(\ref{eq:mf12}) and (\ref{eq:mf12pol}) to the physical pion and
kaon, as follows.
We have first used the experimental
values of the ratios $M_\pi/f_\pi$ and $M_K/M_\pi$ to determine the average
up-down and the strange quark mass respectively. Once these masses have been
determined, we have used again eqs.~(\ref{eq:mf12}) and (\ref{eq:mf12pol}) to compute
the values of the pion and kaon decay constants as well as their ratio
$f_K/f_\pi$.

\section{Physical results}
In order to convert into physical units the results obtained for the strange
quark mass and the kaon decay constants we have fixed the scale within each analysis
(polynomial, PQChPT and C-PQChPT fits) by using $f_\pi$ as physical input. 
In the case of the
polynomial and PQChPT fits we conservatively introduce for the
dimensionful quantities a 6\% and 3\% of systematic error to take into account 
the different scale estimate derived in the analysis over the lightest four
unitary points~\cite{NOI,Boucaud:2007uk}.

The determination of the physical strange and up-down quark masses also requires
implementing a renormalization procedure. The relation between the bare twisted
mass at maximal twist, $\mu_q$, and the renormalized quark mass, $m_q$, is
given by
$m_q(\mu_R) = Z_m(g^2,a \mu_R)\, \mu_q(a)$,
where $\mu_R$ is the renormalization scale, conventionally fixed to 2 GeV for
the light quarks. $Z_m$ is the inverse of the flavour non-singlet pseudoscalar
density renormalization constant, $Z_m=Z_P^{-1}$. We have used the
non-perturbative RI-MOM determination of $Z_P$, which gives
$Z_P^{\rm{RI-MOM}}(1/a) =0.39(1)(2)$ at $\beta=3.9$~\cite{rimom}, and converted
the result to the $\msb$ scheme at the scale $\mu_R=2$ GeV by using
renormalization group improved continuum perturbation theory at the
N$^3$LO~\cite{Chetyrkin:1999pq}.

In table~\ref{tab:physical} we collect the results for the light quark masses
and pseudoscalar decay constants, in physical units, and for the ratios
$m_s/m_{ud}$ and $f_K/f_\pi$, as obtained from the
polynomial, PQChPT and C-PQChPT fits. 
To be conservative, we consider the results obtained from the
analysis of the quark mass combinations satisfying the constraint $\mu_2 \ge
\mu_1 = \mu_S$ which, though being affected by larger statistical errors, are
safe from the effects of the potentially divergent chiral logarithms.
%%%%%%%%%%%%%%%%%%%%%%%%%%%%%%%%%%%%%%%%%%%%%%%%%%%%%%%%%%%%%%%%%%
\begin{table}[t]
\begin{center}
\renewcommand{\arraystretch}{1.4}
\vspace{-0.4cm}
\begin{tabular}{||c||c|c|c|c|c||} \hline 
Fit        & $m_{ud}^{\msb}$ (MeV) & $m_s^{\msb}$ (MeV) & $m_s/m_{ud}$ & $f_K$
(MeV) & $f_K/f_\pi$ \\ \hline

Polynomial & 4.07(9)(33)& 109(2)(9)& 26.7(2)(0)& 158.7(11)(89)& 1.214(8)(0)\\
PQChPT   & 3.82(15)(25)& 107(3)(7)& 27.9(2)(0)& 160.2(15)(54)& 1.225(11)(0)\\ 
C-PQChPT & 3.74(13)(21)& 102(3)(6)& 27.4(3)(0)& 161.8(10)(0)& 1.238(7)(0) \\
\hline
\end{tabular}
\renewcommand{\arraystretch}{1.0}
\end{center}
\vspace{-0.4cm}
\caption{\sl Results for the light quark masses and pseudoscalar decay
constants, in physical units, from the polynomial, PQChPT and
C-PQChPT fits, analysing only the combinations of quark masses
satisfying $\mu_2 \ge \mu_1 = \mu_S$. The quoted errors are
statistical (first) and systematic (second), the latter coming from the
uncertainties in the determination of the lattice scale and of the quark mass
renormalization constant.}
\label{tab:physical}
\vspace*{-0.3cm}
\end{table}
%%%%%%%%%%%%%%%%%%%%%%%%%%%%%%%%%%%%%%%%%%%%%%%%%%%%%%%%%%%%%%%%%%
 In
table~\ref{tab:physical} we quote as a systematic error within each fit the
uncertainty associated with the determination of the lattice spacing and of the
quark mass renormalization constant.

In order to derive our final estimates for the quark masses and decay constants,
we perform a weighted average of the results of the three analyses presented in
table~\ref{tab:physical} and conservatively include the whole spread among
them in the systematic uncertainty. In this way, we obtain as our final
estimates of the light quark masses the results
\beq
\label{eq:mlight}
m_{ud}^{\msb}(2\ \gev)=3.85 \pm 0.12 \pm 0.40 \ \mev  \quad , \quad
m_{s}^{\msb}(2\ \gev)=105 \pm 3 \pm 9 \ \mev \, ,
\eeq
and the ratio 
\beq
m_s/m_{ud}=27.3 \pm 0.3 \pm 1.2\,,
\eeq
where the first error is statistical and the second systematic. For the kaon
decay constant and the ratio $f_K/f_\pi$ we obtain the accurate determinations
\beq
\label{eq:fK}
f_K=161.7 \pm 1.2 \pm 3.1 \ \mev \quad, \quad 
f_K/f_\pi=1.227 \pm 0.009  \pm 0.024 \,.
\eeq

\begin{figure}[!t]
\vspace*{-0.4cm}
\begin{minipage}{16pc}
\includegraphics[scale=0.27,angle=270]{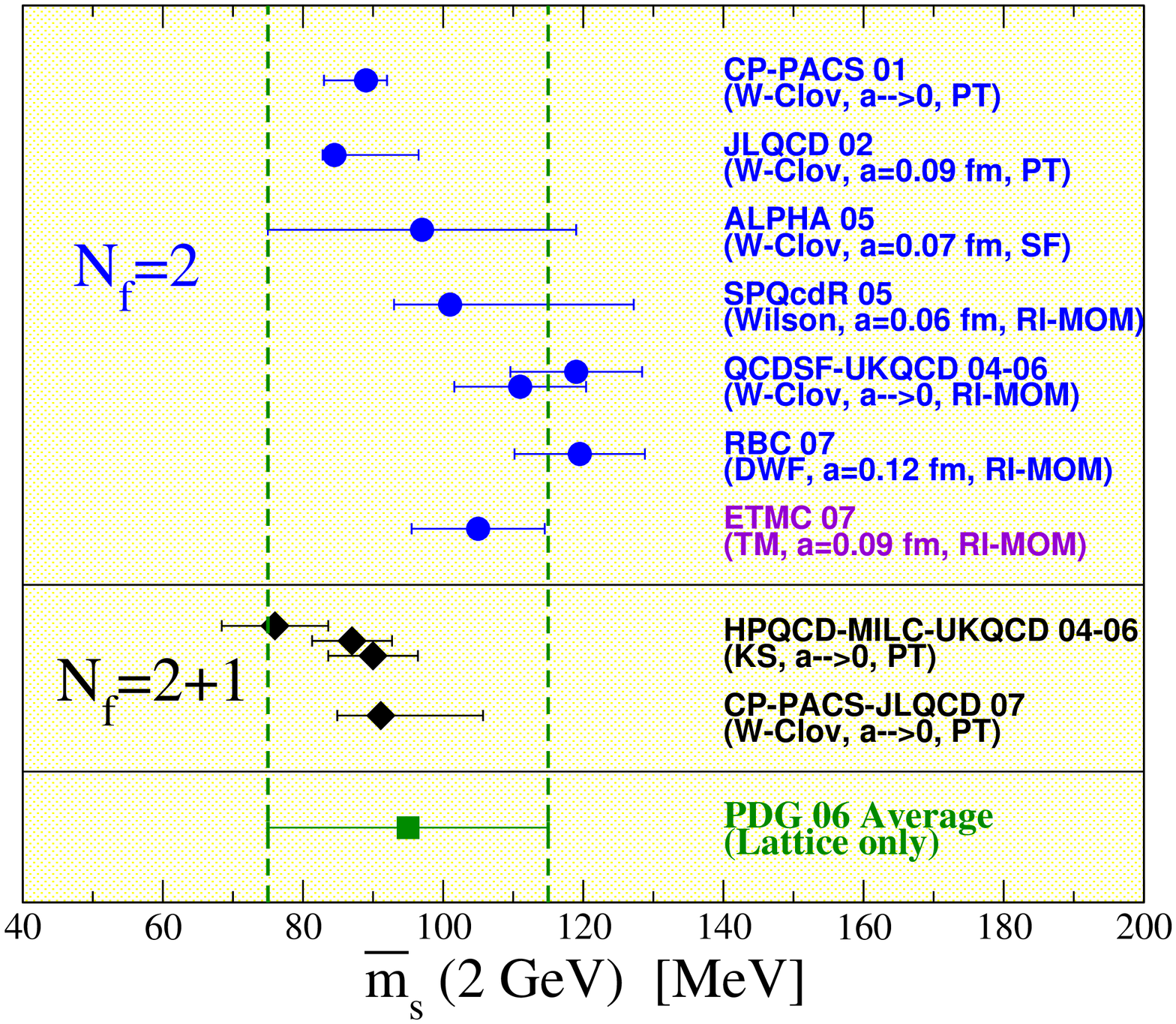} 
\end{minipage}\hspace{2pc}
\begin{minipage}{16pc}
\includegraphics[scale=0.27,angle=270]{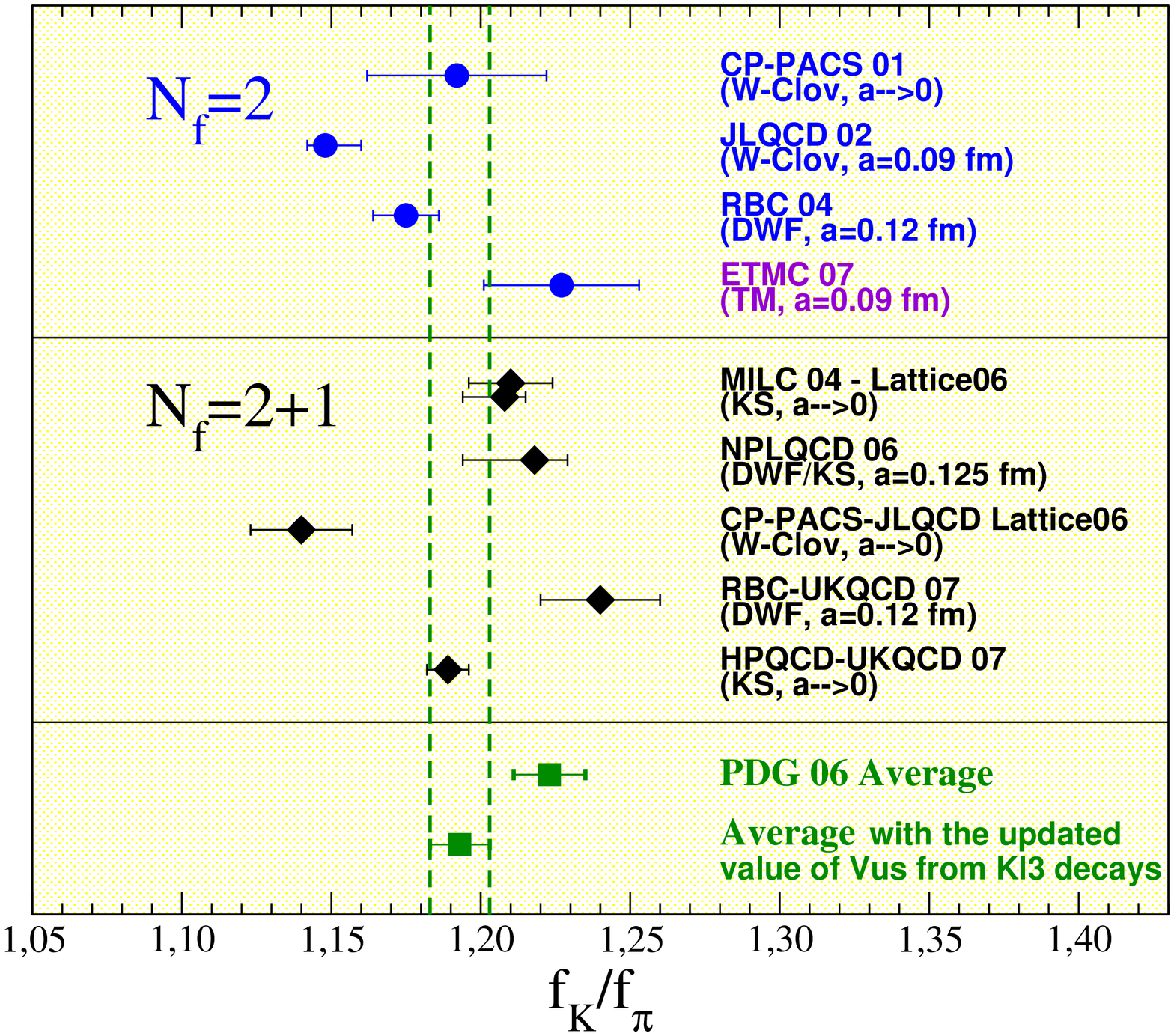}
\end{minipage}
\vspace*{-0.5cm}
\caption{\label{fig:msfkfpiunq} \small\sl Lattice QCD determinations of the
  strange quark mass (left) and of the ratio $f_K/f_\pi$ (right) obtained
from simulations with $N_f=2$ and
$N_f=2+1$
dynamical fermions. The results are also compared with the PDG 06
averages~\cite{pdg} and, for $f_K/f_\pi$, with the average from the $K_{\ell
  3}$ determination of $V_{us}$~\cite{isidori}.}
\vspace*{-0.3cm}
\end{figure}

An interesting comparison of our results for the strange quark mass and the
ratio $f_K/f_\pi$ with other
lattice QCD determinations is illustrated
in fig.~\ref{fig:msfkfpiunq} (see ref.~\cite{NOI} for the
full list of references).

An important finding of our analysis~\cite{NOI} is that the use of non-perturbative
renormalization turns out to play a crucial role in the determination of the quark masses.
The estimate $Z_P^{\rm{RI-MOM}}(1/a)= 0.39(1)(2)$ obtained with the RI-MOM
method is in fact significantly smaller than the prediction $Z_P^{\rm{BPT}}(1/a)
\simeq 0.57(5)$ given by one-loop boosted perturbation theory (in the same
RI-MOM renormalization scheme)~\cite{rimom}. Had we used the perturbative
estimate of $Z_P$ we would have obtained $m_{ud}^{\msb}(2\ \gev)=2.63 \pm 0.08
\pm 0.36 \ \mev$ and $m_{s}^{\msb}(2\ \gev)=72 \pm 2 \pm 9 \ \mev$. As shown in
fig.~\ref{fig:msfkfpiunq} (left), our prediction for the strange quark mass in
eq.~(\ref{eq:mlight}) is in good agreement with other determinations based on a
non-perturbative evaluation of the mass renormalization constant. 
The non-perturbative renormalization method, therefore, is found to have an
important impact that can
be even larger than the quenching effect and that
should be kept in mind, particularly when combining the lattice 
results to  produce the quark mass final averages.
%%%%%%%%%%%%%%%%%%%%%%%%%%%%%%%%%%%%%%%%%%%%%%%%%%%%%%%%%%%%%%%%%%%%%
%%%%%%%%%%%%%%%%%%%%%%%%%%%%%%%%%%%%%%%%%%%%%%%%%%%%%%%%%%%%%%%%%%%%%

Our result for the ratio $f_K/f_\pi$ can be combined with the experimental
measurement of $\Gamma(K \to \mu \bar \nu_\mu (\gamma))/\Gamma(\pi \to \mu \bar
\nu_\mu (\gamma))$~\cite{pdg} to get a determination of the ratio $\vert
V_{us}\vert/\vert V_{ud}\vert$~\cite{marciano}. We obtain
$\vert V_{us}\vert/\vert V_{ud}\vert= 0.2251(5)(47)$,
where the first error is the experimental one and the second is the theory error
coming from the uncertainty on $f_K/f_\pi$. It yields, combined with
the determination $\vert V_{ud} \vert= 0.97377(27)$~\cite{Marciano:2005ec} from
nuclear beta decays, the estimate
$\vert V_{us}\vert= 0.2192(5)(45)$,
in agreement with the value extracted from $K_{\ell 3}$ decays, $\vert V_{us}
\vert= 0.2255(19)$~\cite{isidori}, and leads to the constraint due to the
unitarity of the CKM matrix
$\vert V_{ud}\vert^2 + \vert V_{us}\vert^2 + \vert V_{ub}\vert^2 -1 =
(-3.7 \pm 2.0)\cdot 10^{-3}$.

%In the near future, to further improve the accuracy of the analysis presented 
%here, we plan to
%extend the simulation to other two lattice spacing values (corresponding to
%$\beta=3.8$ and $\beta=4.05$). This should allow us to better quantify the
%size%of discretization effects, of ${\cal O}(a^2)$ in the present
%calculation, and to perform the extrapolation to the continuum limit.

\end{document}